# Simple mathematical model of aging

Morokov Yu. N.

Institute of Computational Technologies SB RAS, Academician M. A. Lavrentiev avenue, 6;
Novosibirsk State University, Pirogova Str., 2,
630090, Novosibirsk, Russia
e-mail: quant0707@gmail.com

**Abstract**

A simple mathematical model of the aging process for long-lived organisms is considered. The key point in this model is the assumption that the body does not have internal clocks that count out the chronological time at scales of decades. At these scales, we may limit ourselves by empirical consideration only the background (smoothed, averaged) processes. The body is dealing with internal biological factors, which can be considered as the biological clocks in suitable parameterization of corresponding variables. The dynamics of these variables is described using a system of autonomous ODEs. A particular representation of the right-hand side of equations in the form of quadratic polynomials is considered. In the simplest case of one variable we deal with a logistic equation, which has an analytical solution. Such quadratic model is justified if it is used to predict the dynamic of aging process for relatively small time intervals. However, since a well-defined biological interpretation can be given for the quadratic right-hand side of the equations, we can expect that the area of applicability of this simplified empirical model can extend to relatively large time intervals. The considered model is, in our opinion, a good and simple mathematical framework for organizing experimental data. The model may be useful to quantify and arrange the chains of causal relationships that may appear significant for the development of the aging process.

**Introduction**

Aging is the result of the destructive impact of metabolic errors and external stress factors on the individual development of the body, expressed in compensatory hyperfunction and failure of systems for maintaining homeostasis (from molecular to organismic levels) and increasing the likelihood of illness and death in life-compatible conditions. The rate of aging in different people of the same age may distinguish significantly. They vary for different systems and organs within the same organism. The aging of one system causes changes in many others (Moskalev, 2019).

There is a very large number of studies on the problem of biological aging. However, the issue of determining the drivers of aging, that is, the main causal factors in the aging process, has been and remains actively discussed (Hayflick, 2007; Kowald, et al., 2016; López-Otín et al., 2016; Pinto da Costa et al., 2016). The problem is that the functioning of a significant part of the biological subsystems changes during aging, but mainly these are concomitant changes that can be considered as just a consequences of the homeostatic reactions of the body to changes in drivers of aging. Such numerous concomitant factors can be useful, for example, as biomarkers of aging.

In this paper, we consider a simple mathematical model, which assumes that at each moment of the chronological time $t$ the state of some biological organism can be described using a finite number $n$ of some variables $g_i(t)$, $i = 1,..., n$. However, within any reasonable model of biological processes, we are forced to limit ourselves to a relative small number $n$, thereby discarding a huge number of variables $g_i(t)$, whose influence on the aging process we decided to neglect.

We will consider aging of long-lived organisms, including humans. Processes of human aging proceed at scales of decades. At these scales, we may limit ourselves by empirical consideration only the background (smoothed, averaged) processes. It is assumed also that at each moment in time the organism is in some kind of quasi-equilibrium state of homeostasis, and, accordingly, any additional "memory effects" are not included, when describing the dynamics of variables $g_i(t)$.

In this work, we do not aim to discuss specific examples of biological variables $g_i(t)$, which could be considered as potential drivers of aging. Such examples are actively discussed in numerous literatures, including those cited by us. We may imagine for concreteness that $g_i(t)$ describe some molecular damages in the body, although this restriction is not fundamental for the model.

## Results and Discussion

We assume that the variables $g_i(t)$, essential for the aging process, are quite smooth functions and satisfy the system of autonomous ODEs:

$$\frac{d\boldsymbol{g}}{dt} = \boldsymbol{f}(\boldsymbol{g}), \tag{1}$$

where $\boldsymbol{g}$ is a vector composed of variables $g_i$.

We can consider the linear transformation of each variable $g_i(t)$

$$\bar{g}_i = \alpha_i g_i + \beta_i, \tag{2}$$

where $\alpha_i$ and $\beta_i$ are constants independent of time $t$. The transformation (2) is a particular case of admissible transformations of the variables $g_i(t)$ retaining the system (1) autonomous. We can always choose the parameters $\alpha_i$ so that each variable $g_i(t)$ will have a dimension of time and, accordingly, it can be interpreted as the corresponding biological time. Thus, in the model with fixed $n$ there will be exactly $n$ biological times.

The main argument in favor of the autonomy of the system (1) is that we do not know any clocks that would physically count out the chronological time in the body at the scale of decades. Therefore, the organism "does not see" the chronological time itself. It is able to "see" only the biological times $g_i(t)$ of the corresponding biological subsystems of a given organism. Accordingly, for the description of the aging process, we assume that the right-hand side of the equation (1) explicitly depends only on $g_i$ and does not depend explicitly on chronological time $t$. It also means that we consider the dynamics of only "natural" physiological aging processes, without any additional "external" targeted interventions (lifestyle change, additional dietary and pharmacological interventions, etc.).

### *The simplest case (n = 1)*

In the case $n = 1$, we are dealing with one scalar variable $g(t)$. The function $g(t)$ is found by solving the Cauchy problem for the equation (1) with the initial condition $g(0) = g_0$. Bearing in mind the possibility of using the transformation (2), we can assume without loss of generality that the variable $g(t)$ increases with time $t$, and since it cannot increase in the body indefinitely, we assume that the organism dies when $g(t)$ reaches some limit value for the survival – $g_{max}$.

Next, we consider a particular representation of $f(g)$ in the form of a quadratic polynomial. Then equation (1) takes the form

$$\frac{dg}{dt} = a + bg + cg^2, \tag{3}$$

where empirical parameters $a$, $b$ and $c$ were entered.

*What is the range of applicability of quadratic representation of f(g)?*

1. We can consider the function $f(g)$ as rather smooth, since we are considering a smooth (background) aging process. It is possible to solve the equation (3) on small time intervals and to redefine the function $g(t)$ on each step using (2), and considering a new $g(t)$ as a small deviation from the obtained value on the previous time step. Here we can consider the right-hand side in (3) as the expansion of the function $f(g)$ into a Taylor series with an accuracy of $O(g^3)$. In this case, the solution of the equation (3) gives some prediction for a small time interval. For a next time step, experimental measurements can be taken anew and the parameters *a, b* and *c* can be identified (refined) again.

2. We can actually give a direct biological sense to the parameters *a*, *b* and *c*. This goes beyond simple formal expansion $f(g)$ into the Taylor series.

The parameter *a* can be interpreted as a constant (time-independent) rate of generation of damages *g* in the organism. The parameter *b* can be associated at $b > 0$ with self-reproduction (e.g., the reproduction of viruses or bacteria) or at $b < 0$ - with the immune response. The parameter *c* can be interpreted at $c > 0$ as accounting (in a linear approximation in *g*) of weakening of the immune system with aging.

Thus, within the model described by equation (3), we have a four-parameter family of virtual aging organisms. Each concrete set of numerical values of the parameters $\{a, b, c, g_0\}$ can be considered as a specific individualization of the organism. The function $g(t)$ corresponding to these parameters gives an individual aging curve of this organism.

*Dependence of individual aging curves on model parameters*

If variable $g(t)$ describes literally some type of damages in the body, then we can additionally assume that $g(t) \geq 0$. Since we want to describe the "natural" aging of the organism, consider the region of 4-dimensional parametric space $\{a, b, c, g_0\}$, limited by inequalities:

$$a > 0, \ c \geq 0, \ g_0 \geq 0. \tag{4}$$

The equation (3) is a well-known logistic equation (e.g., Edwards and Penney, 2004). It has an analytical solution which is convenient to present in the form:

$$g(t) = g_0 + \frac{(g_2 - g_0)(g_1 - g_0)}{(g_2 - g_0) - (g_1 - g_0)e^{-c(g_2 - g_1)t}} \left(1 - e^{-c(g_2 - g_1)t}\right). \tag{5}$$

Here $g_1$ and $g_2$ correspond to stationary states for which the right-hand side in the equation (3) vanishes.

$$g_1 = \frac{1}{2c}\left(-b - \sqrt{b^2 - 4ac}\right), \quad g_2 = \frac{1}{2c}\left(-b + \sqrt{b^2 - 4ac}\right).$$

If $b^2 - 4ac < 0$, then $g_1$ and $g_2$ are complex. The function $g(t)$, of course, remains real. At the same time, the derivative $dg/dt$ does not vanish at any $t > 0$, and moreover $dg/dt > 0$, due to the assumption $a > 0$. That is, $g(t)$ grows monotonously and unlimited with time. This corresponds to unlimited aging.

In the case
$$b^2 - 4ac > 0, \qquad (6)$$
there are two stationary states in which the amount of the damages in the organism is kept constant. These states are obtained under the initial conditions $g_0 = g_1$ or $g_0 = g_2$, with $g_1 < g_2$, and both values of $g_1$ and $g_2$ are either positive (for $b < 0$) or both negative (for $b > 0$).

The case $b > 0$ again corresponds to unlimited aging, since in this case both $g_1$ and $g_2$ are negative. The inequality
$$b < 0 \qquad (7)$$
corresponds to existence of immune response.

So, the possibility of limited aging appears only when both inequalities (6) and (7) are fulfilled simultaneously. In view of (4) this is equivalent to the fulfillment of one inequality
$$b < -2\sqrt{ac}. \qquad (8)$$

An example of a family of integral curves for the equation (3) in the case of the fulfillment of inequality (8) is presented in Figure.

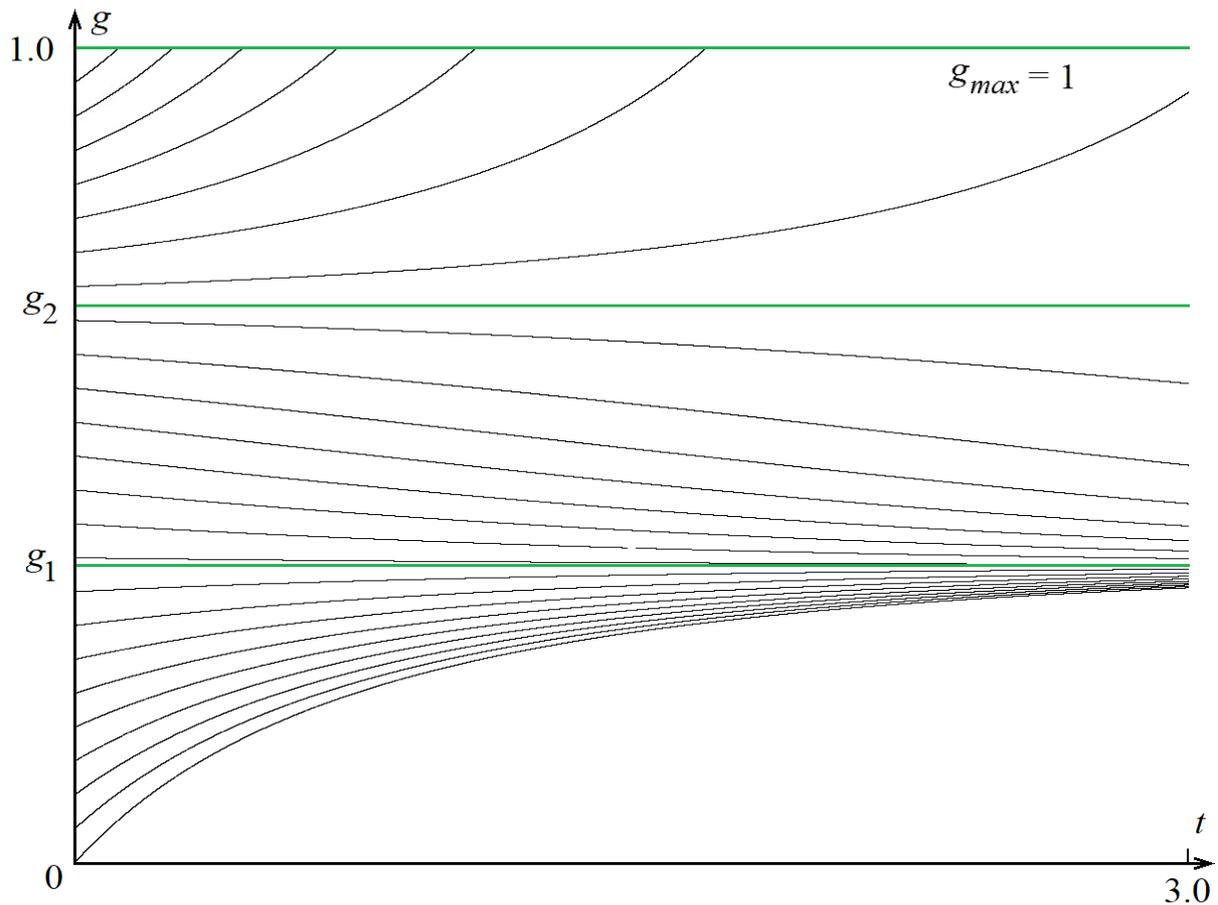

**Figure**. Integral curves for the equation (3) for a set of parameters: $a = 0.5$, $b = -2.1$, $c = 2.0$, $g_{max} = 1$.

The stationary value $g_2$ is unstable. The convergence to the stable stationary value $g_1$ will be only with initial values $g_0 < g_2$.

The function $g(t)$ in (5) has a singularity at time $t_3$, at which the denominator vanishes. This timepoint is determined by the equation

$$e^{-c(g_2-g_1)t_3} = \frac{(g_2-g_0)}{(g_1-g_0)}.$$

The value of the expression in the left part of this relation is in the range from zero to one. Respectively, the same value should take the fraction in the right side. This is achieved only under the initial conditions $g_0 > g_2$. In this case, we have unlimited aging that proceeds faster than exponential, in the sense that $g(t)$ goes to infinity already at a finite value of time $t = t_3$.

Consider the case $c = 0$, when the immunity does not weaken with time. Then the right-hand side of equation (3) will be a linear function of $g$, and

$$g_1 = -\frac{a}{b}, \quad g_2 = +\infty,$$

and

$$g(t) = -\frac{a}{b} + \left(g_0 + \frac{a}{b}\right) \cdot e^{bt}.$$

Here, for positive $b$, $g(t)$ grows unlimited exponentially with time. In the case $b < 0$, there exists a stable positive stationary state $g_1$, to which all integral curves asymptotically converge. The question is only in the ratio between $g_1$ and $g_{max}$. Limited aging will be only for $g_1 < g_{max}$.

In the total absence of immunity, i.e., with $b = c = 0$, we obtain linear unlimited aging for $a > 0$

$$g(t) = g_0 + at.$$

*Multidimensional case* ($n > 1$)

The model described by the equation (3) is naturally generalized to the case of a finite number $n$:

$$\frac{dg_i}{dt} = a_i + \sum_{j=1}^{n} b_{ij} g_j + \sum_{j,k=1}^{n} c_{ijk} g_j g_k. \tag{9}$$

The concrete set of parameters $\{a_i, b_{ij}, c_{ijk}, g_{i0}\}$ used in (9) can be considered as a concrete individualization of the virtual organism. In the approximation, when all parameters $c_{ijk} = 0$, we obtain a system of linear first-order ODEs, the theory of which is well known (e.g., Edwards and Penney, 2004).

$$\frac{dg_i}{dt} = a_i + \sum_{j=1}^{n} b_{ij} g_j. \tag{10}$$

For any concrete realization of the models (9) or (10), we must determine the variables $g_i(t)$ to be included in the model. That is, we must first make a hierarchical list of potential variables, ordering it according to the expected degree of their influence on the aging process. Then we must fix the number $n$, that is, how many variables we include in the model from the top of the list. In any case, it makes sense to start with the simplest models with small $n$.

Above, we have considered in detail the simplest model (3), in which only one variable $g(t)$ is included ($n = 1$). This case can be considered as a special case of the general model (9), in which one variable dominates. The next obvious practical step is the consideration of simplified relations - binary ($n = 2$), triple ($n = 3$), etc.

Perhaps the closest to the considered model is a stochastic dynamical network model of deficits, considered, for example, in (Taneja et al., 2016). Unlike their model we consider, say, the simple model (10) as the simplest empirical scheme that does not contain redundant parameters, but is nevertheless able to quantify the interactions between the variables $g_i(t)$. This allows us to hope to advance in solving the main problem - separating the numerous simply accompanying variables $g_i(t)$ from the drivers of ageing we are looking for.

### *Possible use of the model in solving of three problems: determining the drivers of ageing, finding informative biomarkers of ageing, and determining the biological age*

The most general task of studying the aging process is to determine the drivers of aging, that is, those basic factors that are responsible for the aging process itself. In solving this problem, we initially deal with a huge set $G$ of variables $g_i(t)$, changing over time in the process of ageing.

The same set G is, generally speaking, the initial one when solving two other problems - finding informative biomarkers of aging and determining the biological age (López-Otín et al., 2016; Cohen et al., 2015; Jylhävä et al., 2017; Taneja et al., 2016; Moskalev, 2019).

The main idea of the biomarkers is a quantitative estimation of biological differences in health between people of the same age, do precise and exhaustive appraisal based on numerous analyses and tests' results often appropriate for predictive models' creation. Every quantitative trait of the organism, which is known to change with age, may be used as a biomarker of aging, but consequently not all biomarkers are informative and valuable for diagnostic purposes. The "biological age" is less objective and shows the deficit between predicted individual's life expectancy and the idem marker measured for the cohort or population of the same age. Theoretically and practically many biomarkers may be used for biological age estimation, but the last is depersonalized at least, being connected to the average values known for population and not appropriate for precision medicine (Solovev et al., 2019).

According to (Moskalev, 2019), the most comprehensive online database of human aging biomarkers today is Digital Aging Atlas (http://ageing-map.org). The Digital Ageing Atlas (DAA) is a portal of age-related changes covering different biological levels. It integrates molecular (3071), physiological (343), psychological (17) and pathological (2599) age-related data to create an interactive portal that serves as the first centralised collection of human ageing changes and pathologies (Craig et al., 2015). Here, the numbers of age-related changes of the corresponding type are shown in parentheses.

We cannot be sure that the drivers of aging are already presented in this database. Nevertheless, we can set a task - the selection of the most significant candidates for the drivers of aging from this particular database. That is, we must discard all the numerous concomitant factors $g_i(t)$. This problem could be solved purely technically if, say, we had a set of identified empirical parameters $\{a_i, b_{ij}\}$ in the system of equations (10).

To discard a significant part of the concomitant quantities $g_i(t)$ from this database, it is not necessary to consider all the parameters $\{a_i, b_{ij}\}$. Many of these concomitant quantities could most likely be discarded at the stage of consideration of simplified relations - binary ($n = 2$), triple ($n = 3$), etc.

The problem of determining informative biomarkers of aging also solves the problem of discarding uninformative concomitant quantities $g_i(t)$, but the selection criteria differ significantly from the task of determining the drivers of aging.

For example, the following main criteria for aging biomarker were proposed (Butler et al., 2004; Moskalev, 2019):

• Must change with age;
• Have to predict mortality better than chronological age;
• Allow foreseeing the early stages of a specific age-related disease;
• To be minimally invasive - do not require serious intervention or painful procedure.

The redundancy of diverse data creates a great difficulty, so called "curse of multidimensionality", in analysis and interpretation of the results when a complex study have been carrying out during which the thousands of dimensions (measurements of biomarkers' values) appear. A computational approach engaging deep machine-learning algorithms and neural networks in biomarker selection opens a new perspective for systemic analysis of populations' big data and for population specific biomarker panels' creation (Solovev et al., 2019).

## Conclusion

Here we have considered a simple mathematical model of the aging process, based on a system of first-order ODEs. The key point in this aging model is the assumption that the body does not have an internal clock that counts out the chronological time at the scale of decades. The body is dealing with internal biological factors, which can be considered as the biological clocks. We associate with these factors the variables $g_i(t)$. Each variable $g_i$ counts out, in suitable parameterization, its own biological time $t_i$. This assumption enables us to consider the system (1) as the system of autonomous ODEs applied to the description of the aging process.

The model is very simple and the use of ODE is the standard when considering various biological processes. Writing the functions $f_i(g)$ on the right-hand side of the equations (3) and (9) in the form of quadratic polynomials is just a good approximation and, in general, is not necessary. However, such writing admits an explicit analytical solution in the one-dimensional case ($n = 1$) and in the linear multidimensional case (10). Progress in developing of accuracy of the model is determined by progress in obtaining reliable experimental data that could help identify the parameters at the right side of the equations (3), (9) or (10).

In this work, we focused only on the model of the aging process. Mortality, closely related to the ageing process, was considered symbolically only in the simplest way - through the limit value $g_{max}$. However, the consideration of the mortality problems is beyond the scope of this work.

Overall, the considered simple mathematical model is, in our opinion, a good and simple mathematical framework for organizing experimental data on changes in various biological subsystems of the body with age. Even in the linear approximation by $g_i$ at the right side of equations (10), the parameters $\{a_i, b_{ij}\}$ allow to quantify and arrange the chains of causal relationships that may appear essential for the development of the ageing process.